\documentclass[11pt,reqno]{amsart}
\usepackage{amsfonts}
\usepackage{mathrsfs}
\usepackage{amsmath}
\usepackage{amssymb,amsfonts}

\newtheorem{thm}{Theorem}[section]
\newtheorem{lem}{Lemma}[section]
\newtheorem{defn}{Definition}[section]
\newtheorem{prop}{Proposition}[section]
\newtheorem{coro}{Corollary}[section]\numberwithin{equation}{section}
\newtheorem{rmk}{Remark}[section]

\def\pf{{\textit {Proof:} }}

\newcommand{\mysection}[1]{\section{#1}\setcounter{equation}{0}}

\newfont{\bb}{msbm10 at 12pt}

\def\hK{\hbox{\bb K}}
\def\R{\hbox{\bb R}}
\def\hC{\hbox{\bb C}}

\def\B{\mathcal B}
\def\C{\mathcal C}
\def\D{\mathcal D}

\def\K{\mathcal K}

\newcommand{\bal}{\begin{aligned}}      \newcommand{\eal}{\end{aligned}}
\newcommand{\ba}{\begin{array}}      \newcommand{\ea}{\end{array}}
\newcommand{\bc}{\begin{center}}     \newcommand{\ec}{\end{center}}
\newcommand{\be}{\begin{enumerate}}  \newcommand{\ee}{\end{enumerate}}
\newcommand{\beq}{\begin{eqnarray}}  \newcommand{\eeq}{\end{eqnarray}}
\newcommand{\beQ}{\begin{eqnarray*}} \newcommand{\eeQ}{\end{eqnarray*}}
\newcommand{\bi}{\begin{itemize}}    \newcommand{\ei}{\end{itemize}}
\newcommand{\bt}{\begin{tabular}}    \newcommand{\et}{\end{tabular}}
\newcommand{\bdm}{\begin{displaymath}} \newcommand{\edm}{\end{displaymath}}

\def\qed{\hfill{Q.E.D.}\smallskip}

\newcommand{\ls}{\setlength{\baselineskip}{12pt}
                 \setlength{\parskip}{3mm}}

\begin{document}

\title[]{The positive energy theorem for asymptotically anti-de Sitter spacetimes}

\author{Yaohua Wang}
\address[Yaohua Wang]{School of Mathematics and Statistics, Henan University, Kaifeng, Henan 475004, PR China}
\email{wangyaohua@henu.edu.cn}
\author{Naqing Xie}
\address[Naqing Xie]{Institute of Mathematics, School of Mathematical Sciences, Fudan
University, Shanghai 200433, PR China}
\email{nqxie@fudan.edu.cn}
\author{Xiao Zhang}
\address[Xiao Zhang]{Institute of Mathematics, Academy of Mathematics and
Systems Science, Chinese Academy of Sciences, Beijing 100190, PR China}
\email{xzhang@amss.ac.cn}

\date{}

\begin{abstract}

We establish the inequality for Henneaux-Teitelboim's total energy-momentum for asymptotically anti-de Sitter initial data
sets which are asymptotic to arbitrary $t$-slice in anti-de Sitter spacetime. In particular, when $t=0$, it generalizes
Chru\'{s}ciel-Maerten-Tod's inequality in the center of AdS mass coordinates. We also show that the determinant of energy-momentum
endomorphism ${\bf Q}$ is the geometric invariant of asymptotically anti-de Sitter spacetimes.\\\\
Keywords: General relativity; the positive energy theorem; asymptotically anti-de Sitter spacetime\\
Mathematics Subject Classification 2010: 53C27; 53C80; 83C4

\end{abstract}

\maketitle \pagenumbering{arabic}

\mysection{Introduction}\ls

The positive energy theorem plays a fundamental role in general relativity. When the cosmological constant is zero and spacetimes are
asymptotically flat, the positive energy theorem for the ADM total energy-momentum \cite{ADM} was first proved by Schoen and Yau \cite{SY1,SY2,SY3},
then by Witten \cite{Wi,PT}. We refer to \cite{EHLS, B, D, XD} for the case of higher dimensional spacetimes.

When the cosmological constant is negative and spacetimes are asymptotically anti-de Sitter, initial data sets are asymptotically hyperbolic and the second fundamental forms are asymptotic to zero. There are a large number of papers to devote to define the total energy-momentum and prove its positivity in a physical manner, see, e.g. \cite{AD, HT, ACOTZ} and references therein. (It seems the total energy was first defined in \cite{AD}, and which also contained the proof of its positivity via SUGRA, exactly as the proof for zero cosmological constant \cite{DT}.) However, the mathematical rigorous and complete proofs were given only in \cite{Wa, CH} for asymptotically anti-de Sitter initial data sets with zero second fundamental form, and in \cite{M, CMT} for the initial data sets with nontrivial second fundamental form where the energy-momentum matrix was proved to be positive semi-definite. And some energy-momentum inequalities were
proved with respect to certain specific coordinate systems in \cite{CMT}.

There is also another version of the positive energy theorem for asymptotically hyperbolic manifolds \cite{Z, CJL, XZ} representing initial data sets near null infinity in asymptotically flat spacetimes. In this case both the metrics and the second fundamental forms are asymptotic to the hyperbolic metric. In particular, the theorem in \cite{Z, XZ} gives a different energy-momentum inequality for asymptotically anti-de Sitter initial data sets with the nontrivial second fundamental form if its trace is nonpositive.

The anti-de Sitter spacetime can be viewed as the hyperboloid
\beq
\eta_{\alpha\beta}y^\alpha y^\beta=\frac{3}{\Lambda}, \qquad \Lambda =-3 \kappa ^2 \ (\kappa >0) \label{ads}
\eeq
in $\R^{3,2}$ equipped with the metric
\beQ
\eta_{\alpha\beta}dy^\alpha dy^\beta= -(dy^0)^2+\sum^3_{i=1}(dy^i)^2-(dy^4)^2.
\eeQ
There are ten Killing vectors generating rotations for $\R^{3,2}$
\beq
U_{\alpha\beta}=y_\alpha \frac{\partial}{\partial y^\beta}-y_\beta\frac{\partial}{\partial y^\alpha}.  \label{ten-K}
\eeq
Under coordinate transformations
\beq\label{y}
y^0=\frac{\cos(\kappa t)}{\kappa} \cosh(\kappa r),\
y^i=\frac{1}{\kappa}\sinh(\kappa r) n^i,\
y^4=\frac{\sin(\kappa t)}{\kappa} \cosh(\kappa r),
\eeq
where $n^1=\sin\theta\cos\psi $, $n^2=\sin\theta\sin\psi$, $n^3=\cos \theta$, the induced anti-de Sitter metric is
\begin{equation}
\widetilde{g}_{AdS}=-\cosh^2(\kappa r)dt^2+dr^2+\frac{\sinh^2(\kappa r)}{\kappa^2}\big(d\theta^2+\sin^2\theta d \psi^2\big).
\label{ads-metric}
\end{equation}
Let the coframe of (\ref{ads-metric}) be
\beQ
\breve{e}^0=\cosh(\kappa r)dt,\ \breve{e}^1=dr, \ \breve{e}^2=\frac{\sinh(\kappa r)}{\kappa}d\theta, \
\breve{e}^3=\frac{\sinh(\kappa r)\sin\theta}{\kappa} d \psi
\eeQ
and denote $\{\breve{e}_\alpha \}$ as its dual frame.

The metric and the second fundamental form of $t$-slice are the same in
(\ref{ads-metric}) no matter that $t=0$ or not. However, $U_{\alpha\beta}$ are different restricting on different $t$-slices and depend on $t$
(cf. Appendix A). In \cite{HT}, Henneaux and Teitelboim defined the total energy-momentum for asymptotically anti-de Sitter spacetimes
\beq
J_{ab} ^{HT} =\lim_{r\rightarrow \infty}\int_{S_r}{\breve{G}^{ijkl} [ U_{ab}^\perp \breve{\nabla}_j g_{kl}-\breve{\nabla}_j U_{ab}^\perp  a_{kl}]}dS_i+\lim_{r\rightarrow \infty}\int_{S_r}2 U_{ab}^{(k)} \pi_{k}^{\ i}dS_i, \label{HT-em}
\eeq
where initial data sets ($M$, $g$, $h$) are asymptotic to $t$-slice of (\ref{ads-metric}), $a_{kl}=g_{kl}-\breve{g}_{kl}$, and $\breve{g}$,
$\breve{\nabla}$ are the metric and the Levi-Civita connection of $t$-slice of (\ref{ads-metric}) respectively,
\beQ
\breve{G}^{ijkl}=\frac{1}{2}\sqrt{\breve{g}}\big(\breve{g}^{ik}\breve{g}^{jl}+\breve{g}^{il}\breve{g}^{jk}-2\breve{g}^{ij}\breve{g}^{kl}\big),
\quad \pi_{k}^{\ i} =h_{k}^{i}-\delta _{ki} tr_{\breve{g}}(h).
\eeQ
These quantities form an energy-momentum endomorphism ${\bf Q}$. When $t=0$, (\ref{HT-em}) reduce to the definitions provided in \cite{Wa, CH, M, CMT}.

Recall that, using essentially the explicit forms of $U_{\alpha \beta}$ for $t=0$, Chru\'{s}ciel, Maerten and Tod \cite{CMT} provided definitions of the total
energy $m_{(\nu)}\  (\nu=0,1,2,3)$, the rest-frame angular momentum $j_{(i)}$ and the center of mass $c_{(i)}$ ($i=1,2,3$),
with respect to the anti-de Sitter spacetime equipped with the metric
\beQ
\widetilde{g}_{AdS}=-\Big(\frac{1+|x|^2}{1-|x|^2}\Big)^2dt^2+\frac{4}{(1-|x|^2)^2}\sum^3_{i=1}(dx^i)^2.
\eeQ
Denote by $\nabla$ and $\breve{\nabla}$ the Levi-Civita connections of the initial data sets with respect to the metric $g$ and the background hyperbolic metric
$\breve{g}$ respectively. The total energy vector  $m_{(\nu)}\  (\nu=0,1,2,3)$ is defined as
\beQ
m_{(\mu)}=\frac{1}{8\pi}\lim_{|x|\rightarrow 1^-}\int_{S_{|x|}}\sqrt{\det g} \big(V_{(\mu)}g^{i[k} g^{j]l} \breve{\nabla}_j g_{kl}
+\nabla^{[i}V_{(\mu)}g^{j]k} (g_{jk}-\breve{g}_{jk})\big) dS_i
\eeQ
where $V_{(0)}=\frac{1+|x|^2}{1-|x|^2}$, $V_{(j)}=\frac{(-2)x^j}{1-|x|^2}$. Let $Y$ be a tangential vector to the $t=0$ slice. Denote
\beQ
H(Y)=\frac{1}{8\pi}\lim_{|x|\rightarrow 1^-}\int_{S_{|x|}}\sqrt{\det g}\big(h^i{_j}-h^k{_k}\delta^i_j\big)Y^j dS_i,
\eeQ
where $h_{ij}$ is the second fundamental form of the slice in the spacetime. The rest-frame angular momentum vector $j_{(i)}$ ($i=1,2,3$) is
\beQ
j_{(i)}=\epsilon_{ijl}H(\Omega_{(j)(l)})
\eeQ
where
$\Omega_{(j)(l)}=x_j\partial_l-x_l\partial_j$. And the center of mass vector $c_{(i)}$ is
\beQ
c_{(i)}=H(C_{(i)})\eeQ
where $C_{(l)}=\big(\frac{1+|x|^2}2\delta^j _l -
 {x^j x^l}\big)\partial_j$. Denote
\beQ
\vec{m}=(m_{(1)}, m_{(2)}, m_{(3)}), \quad \vec{c}=(c_{(1)}, c_{(2)}, c_{(3)}), \quad \vec{j}=(j_{(1)}, j_{(2)}, j_{(3)}).
\eeQ
They pointed out that \cite{CMT}, if the total energy 4-vector is timelike, i.e.,
\beQ
m_{(0)}> \big(m_{(1)}^2 +m_{(2)}^2 +m_{(3)}^2 \big)^{\frac{1}{2}},
\eeQ
one can make $SO(3,1)$ coordinate transformations such that
\beQ
\begin{aligned}
\big(m_{(0)}^2-m_{(1)}^2 - m_{(2)}^2 - m_{(3)}^2 \big)^{\frac{1}{2}} &\longrightarrow m_{(0)}, \\
m_{(1)}, m_{(2)}, m_{(3)}, c_{(2)}, j_{(1)}, j_{(2)} &\longrightarrow 0,
\end{aligned}
\eeQ
and they proved the energy-momentum inequality
\begin{equation}\label{i}
 m_{(0)}\geq \sqrt{| \vec{c}|^2+| \vec{j} |^2+ 2| \vec{c}\times \vec{j} |}
\end{equation}
in this new coordinate system. We refer to the coordinates satisfying
\beq
m_{(1)}=m_{(2)}=m_{(3)}=c_{(2)}=j_{(1)}=j_{(2)}=0 \label{cond-CMT}
\eeq
as the ``center of AdS mass" coordinates (cf. Appendix B).

Indeed, Witten's argument indicates that ${\bf Q}$ is positive semidefinite. But it does not give that the total energy 4-vector is timelike
for general nontrivial initial data sets (cf. Remark \ref{positive-energy4}). Also the form of (\ref{i}) is not $SO(3,1)$ invariant, and it
changes when it is transformed back to the non-center of AdS mass coordinates. These motivate us to establish the inequality for Henneaux and
Teitelboim's total energy-momentum in general non-center of AdS mass coordinates. In this paper, we prove (Theorem \ref{pmt})
 \beQ
E_0 \geq \sqrt{L^2 -2V^2 +2 \big(\max\{A^4 -L^2 V^2, 0\}\big)^\frac{1}{2} }.
 \eeQ
(See (\ref{e-m}), (\ref{3constants}) for the definitions of these notations.) If three vectors ${\bf c}$, ${\bf c}'$, ${\bf J}$ or $\vec{m}$,
$\vec{c}$, $\vec{j}$ are linearly dependent, i.e, $V=0$, then
\beQ
E_0 \geq \sqrt{L^2 +2 A^2}.
\eeQ
This generalizes the energy-momentum inequality (\ref{i}).

We remark that, unlike the case of non-positive cosmological constant where it always holds and serves as the feature
of spacetimes, the positive energy theorem for the positive cosmological constant holds only on certain very restricted spacelike
hypersurfaces \cite{LXZ, LZ}.

The paper is organized as follows:
In Section 2, we discuss the relation of the total energy-momenta given in \cite{HT} and \cite{CMT}.
In Section 3, we define the energy-momentum endomorphism ${\bf Q}$ and compute it explicitly under a fixed Clifford multiplication.
In Section 4, we establish the new inequality for Henneaux and Teitelboim's total energy-momentum.
In Section 5, we show that ${\bf Q}$ is the geometric invariant of asymptotically anti-de Sitter spacetimes.
In Appendix A, we provide the restriction of the ten Killing vectors $U_{\alpha \beta}$ on the anti-de Sitter spacetime.
In Appendix B, we explicitly construct the center of AdS mass coordinate transformations on the $t=0$ slice.
In Appendix C, we provide roots of the determinant of ${\bf Q}$.

Throughout the paper, repeating indices
means taking summation, with Greek indices running from $0$ to $3$, the lower-case Latin indices running from $1$ to $3$ and
upper-case Latin indices running from $1$ to $2$.

\mysection{Total energy-momentum}\ls

Let $(N,\widetilde{g})$ be a spacetime with negative cosmological constant $\Lambda$, and $\widetilde{g}$
satisfies the Einstein field equations
\begin{equation}
\widetilde{Ric}-\frac{\widetilde{R}}{2}\widetilde{g}+\Lambda \widetilde{g}=T. \label{EinsteinEqs}
\end{equation}
Suppose that the stress-energy tensor $T$ satisfies the dominant energy condition
\begin{equation}
 T_{00}\geq \sqrt{\sum_iT_{0i}^2} ,\ \ T_{00}\geq |T_{\alpha\beta}|. \label{dec}
\end{equation}

Let $(M, g, h)$ be an initial data set where $M$ is a 3-dimensional spacelike hypersurface with the induced
Riemannian metric $g$ and the second fundamental form $h$. Let $\{\breve{e}_i \}$ be the frame of (\ref{ads-metric}).
Recall $\kappa =\sqrt{-\frac{\Lambda}{3}}$. $(M,g,h)$ is said to be {\em asymptotically anti-de Sitter} of order $\tau >\frac{3}{2}$ if\\
(1) There is a compact set $K  \subset M$ such that $M
\setminus K$ is the disjoint union of a finite number of subsets (ends) $M_i$ and each $M_i$
is diffeomorphic to $\mathbb{R}^3 \setminus B_r$ with $B _r$ the closed ball of radius $r$;\\
(2) Under this diffeomorphism, the metric
$g_{ij}=g(\breve{e}_i,\breve{e}_j)$ on each end
is of the form $g_{ij}=\delta_{ij}+a_{ij}$ where $a_{ij}$ satisfies
\beQ
a_{ij}=O(e^{-\tau \kappa r}), \ \breve{{\nabla}}_k a_{ij}=O(e^{-\tau \kappa r}), \ \breve{\nabla}_l\breve{{\nabla}}_k a_{ij}=O(e^{-\tau \kappa r});
\eeQ
and the second fundamental form $h_{ij}=h(\breve{e} _i,\breve{e} _j)$ satisfies
\beQ
h_{ij}=O(e^{- \tau \kappa r}), \ \breve{{\nabla}}_k h_{ij}=O(e^{-\tau \kappa r});
\eeQ
(3) There exists a distance function $\rho _z$ such that $T_{00} e^{\kappa \rho_z}$, $T_{0i} e^{\kappa \rho_z}$ $\in L^1(M)$.
Here $\breve{{\nabla}}$, $\{\breve{e}_i\}$ are the Levi-Civita connection and frame of the hyperbolic metric
\beQ
\breve{g}=dr^2+\frac{\sinh^2(\kappa r)}{\kappa^2}\big(d\theta^2+\sin^2\theta d \psi^2\big)
\eeQ
respectively. Denote
\beQ
\mathcal{E}_i=\breve{\nabla}^j g_{ij}-\breve
{{\nabla}}_itr_{\breve{g}}(g)-\kappa(a_{1i}-g_{1i}tr_{\breve{g}}(a)),\quad
\mathcal{P}_{ki}=h_{ki}-g_{ki}tr_{\breve{g}}(h).
\eeQ
Let $U_{\alpha \beta}$ be the restrictions of the Killing vectors (\ref{ten-K}) on the $t$-slice.
For the convenience of the statement of our main theorem, we introduce the following notions.
\beq\label{e-m}
\begin{aligned}
E_0=&\frac{\kappa}{16\pi}\lim_{r\rightarrow \infty}\int_{S_r}\mathcal{E}_1 U_{40}^{(0)}\breve{\omega},\\
c_{i}=&\frac{\kappa}{16\pi}\lim_{r\rightarrow \infty}\int_{S_r}\mathcal{E}_1 U_{i4}^{(0)}\breve{\omega}
       +\frac{\kappa}{8\pi}\sum_{j=2}^{3}\lim_{r\rightarrow \infty}
       \int_{S_r}\mathcal{P}_{j1}U_{i4}^{(j)} \breve{\omega},\\
c'_{i}=&\frac{\kappa}{16\pi}\lim_{r\rightarrow \infty}\int_{S_r}\mathcal{E}_1 U_{i0}^{(0)}\breve{\omega}
        +\frac{\kappa}{8\pi}\sum_{j=2}^{3}\lim_{r\rightarrow\infty}
        \int_{S_r}\mathcal{P}_{j1}U_{i0}^{(j)} \breve{\omega},\\
J_{i}=&\frac{\kappa}{8\pi}\sum_{j=2}^{3}\lim_{r\rightarrow \infty}
       \int_{S_r}\mathcal{P}_{j1}V_{i}^{(j)} \breve{\omega}, \qquad J_{jl}= \varepsilon _{ijl} J_i
\end{aligned}
\eeq
where $\breve{\omega}= \breve{e}^2\wedge \breve{e}^3$, $U_{\alpha\beta}=U_{\alpha\beta}^{(\gamma)}\breve{e}_{\gamma}$,
$\varepsilon _{ijl} V_i=U_{jl}$. In the frame of (\ref{ads-metric}),
\beQ
J_{ab} ^{HT} =\lim_{r\rightarrow \infty}\int_{S_r}{\breve{G}^{1jkl} [ U_{ab}^{(0)} \breve{\nabla}_j g_{kl}-\breve{\nabla}_j U_{ab}^{(0)} a_{kl}]}\breve{\omega}
+\lim_{r\rightarrow \infty}\int_{S_r}2 U_{ab}^{(k)} \mathcal{P}_{k1}\breve{\omega}.
\eeQ
Since
\beQ
\begin{aligned}
\breve{G}^{1jkl} \big[ & U_{ab}^{(0)} \breve{\nabla}_j g_{kl}-\breve{\nabla}_j U_{ab}^{(0)} a_{kl}\big]\\
=& \frac{1}{2}\big(\delta^{1k}\delta^{jl}+\delta^{1l}\delta^{jk}-2\delta^{1j}\delta^{kl}\big)
   \big[ U_{ab}^{(0)} \breve{\nabla}_j g_{kl}-\breve{\nabla}_j U_{ab}^{(0)} a_{kl}\big]\\
=& \delta^{1k}\delta^{jl}\big[ U_{ab}^{(0)} \breve{\nabla}_j g_{kl}-\breve{\nabla}_j U_{ab}^{(0)} a_{kl}\big]
   -\delta^{1j}\delta^{kl}\big[ U_{ab}^{(0)} \breve{\nabla}_j g_{kl}-\breve{\nabla}_j U_{ab}^{(0)} a_{kl}\big]\\
=& \big(U_{ab}^{(0)} \breve{\nabla}^j g_{1j}-\breve{\nabla}^j U_{ab}^{(0)} a_{1j}\big)
   -\big(U_{ab}^{(0)} \breve{\nabla}_1 tr_{\breve{g}}(g)-\breve{\nabla}_1 U_{ab}^{(0)} tr_{\breve{g}}(a)\big)\\
=& \big(U_{ab}^{(0)} \breve{\nabla}^j g_{1j}-\kappa U_{ab}^{(0)} a_{11}\big)-
   \big(U_{ab}^{(0)} \breve{\nabla}_1 tr_{\breve{g}}(g)-\kappa U_{ab}^{(0)} tr_{\breve{g}}(a)\big)+o(e^{-2\kappa r})\\
=& \mathcal{E}_1U_{ab}^{(0)}+o(e^{-2\kappa r}),
\end{aligned}
\eeQ
we obtain
\beq
J_{40} ^{HT}=\frac{ 16\pi}{\kappa}E_0,\ \
J_{i4} ^{HT}=\frac{ 16\pi}{\kappa}c_{i},\ \
J_{i0} ^{HT}=\frac{ 16\pi}{\kappa}c'_{i},\ \
J_{jl} ^{HT}=\frac{ 16\pi}{\kappa} \varepsilon_{ijl}J_{i} \label{em-and-HT}
\eeq
where $J_{ab} ^{HT}$ is Henneaux-Teitelboim's total energy-momentum (\ref{HT-em}).

Now we discuss the relationship between the quantities (\ref{e-m}) and the total energy-momentum defined in \cite{CMT}.
The original definition is given for $\kappa=1$. But we consider the general $\kappa$ in the followings.
The transformations connecting the hyperbolic metric
$b=\frac{4}{\kappa ^2 (1-|x|^2)^2}dx^2$ used in \cite{CMT} and the metric $\breve{g}$ used in our setting are
\beQ
x^1=\tanh{\frac{\kappa r}{2}}\sin\theta \cos\psi,\ \  x^2=\tanh{\frac{\kappa r}{2}}\sin\theta \sin\psi, \ \
x^3=\tanh{\frac{\kappa r}{2}}\cos\theta.
 \eeQ
Straightforward computation yields
 \beQ
 \begin{aligned}
\partial_{x^1}&=\frac{2}{\kappa}\cosh^2{\frac{\kappa r}{2}}\sin\theta \cos\psi{\partial_r}
                               +\frac{\cos\theta\cos\psi}{\tanh{\frac{\kappa r}{2}}}{\partial _\theta}
                               -\frac{\sin\psi}{\tanh{\frac{\kappa r}{2}}\sin\theta}{\partial _\psi},\\
\partial_{x^2}&=\frac{2}{\kappa}\cosh^2{\frac{\kappa r}{2}}\sin\theta \sin\psi {\partial_r}
                               +\frac{\cos\theta\sin\psi}{\tanh{\frac{\kappa r}{2}}}{\partial_\theta}
                               +\frac{\cos\psi}{\tanh{\frac{\kappa r}{2}}\sin\theta}{\partial_\psi},\\
\partial_{x^3}&=\frac{2}{\kappa}\cosh^2{\frac{\kappa r}{2}}\cos\theta{\partial _r}
                               -\frac{\sin\theta}{\tanh{\frac{\kappa r}{2}}}{\partial _\theta}.
 \end{aligned}
 \eeQ
Thus, in the polar coordinates, the vectors used in \cite{CMT} are
 \beQ
 \begin{aligned}
V_{(0)}&=\cosh \kappa r,  \\
V_{(1)}&=-\sinh\kappa  r\sin\theta \cos\psi,\\
V_{(2)}&=-\sinh\kappa  r\sin\theta \sin\psi,  \\
V_{(3)}&=-\sinh \kappa r \cos\theta,\\
C_{(1)}&=\coth{\kappa r}\Big(\cos\theta\cos\psi{\partial_\theta}
         -\frac{\sin\psi}{\sin\theta}{\partial _\psi}\Big)
         +\frac{1}{\kappa}\sin\theta \cos\psi{\partial _r},\\
C_{(2)}&=\coth{\kappa r}\Big(\cos\theta\sin\psi{\partial_\theta}
         +\frac{\cos\psi}{\sin\theta}{\partial _\psi}\Big)
         +\frac{1}{\kappa}\sin\theta \sin\psi{\partial _r},\\
C_{(3)}&=-\coth{\kappa r}\sin\theta{\partial _\theta}+\frac{1}{\kappa}\cos\theta{\partial _r},\\
\Omega_{(1)(2)}&={\partial _\psi},\\
\Omega_{(2)(3)}&=-\sin\psi{\partial _\theta}
                 -\frac{\cos\theta \cos \psi}{\sin\theta}{\partial _\psi},\\
\Omega_{(3)(1)}&=\cos\psi{\partial _\theta}
                 -\frac{\cos\theta\sin\psi}{\sin \theta}{\partial _\psi}.
 \end{aligned}
 \eeQ

\begin{prop}
The following relations hold between Henneaux-Teitelboim's total energy-momentum and Chru\'{s}ciel-Maerten-Tod's total energy-momentum
  \beq
  \begin{aligned}
E_0 &=m_{(0)}, \\
c_i &=-m _{(i)} \cos \kappa t + c _{(i)} \sin \kappa t,\\
c' _i &= m _{(i)} \sin \kappa t + c _{(i)} \cos \kappa t,\\
J_l &=j_{(l)}. \label{relations}
  \end{aligned}
  \eeq
\end{prop}
\pf By the explicit expressions of $U_{\alpha \beta}$ in Appendix A, we find that $E_0$, $J_i$ do not depend on $t$, and
\beQ
\frac{dc_i}{dt}=\kappa c' _i,\quad \frac{dc' _i}{dt}=-\kappa c _i.
\eeQ
Note the total energy-momentum in \cite{CMT} is defined on $t=0$ slice. And straightforward computation shows that, at $t=0$,
\beQ
E_0 =m_{(0)}, \quad c_i =-m _{(i)}, \quad c' _i = c _{(i)}, \quad  J_l =j_{(l)}.
\eeQ
This yields (\ref{relations}). \qed

In \cite{C}, Carter obtained a family of solutions for the Einstein field equations.
\beQ
ds^2=\frac{\Delta_{\mu}(d\chi-\lambda^2d\psi)^2- \Delta_{\lambda}(d\chi+\mu^2d\psi)^2}{\lambda^2+\mu^2}
+(\lambda^2+\mu^2)\Big(\frac{d\lambda^2}{\Delta_{\lambda}} +\frac{d\mu^2}{\Delta_{\mu}}\Big),
\eeQ
where
\beQ
\begin{aligned}
\Delta_{\lambda}=\frac{1}{3}\Lambda \lambda^4+h\lambda^2-2m\lambda+p+e^2,\quad
\Delta_{\mu}=\frac{1}{3}\Lambda \mu^4-h\mu^2+2q\mu+p
\end{aligned}
\eeQ
and $-\infty <\chi,\ \lambda, \ \mu <\infty$, $0 \leq \psi < 2\pi$.
It provides the Kerr-anti-de Sitter spacetimes if $\Delta_{\lambda}$,  $\Delta_{\mu}$ are given as follows
\beQ
\begin{aligned}
\Delta_{\lambda}=\big(\kappa^2 \lambda^2+1\big)\big(\lambda^2+a^2\big)-2m\lambda,\quad
\Delta_{\mu}=\big(\kappa^2 \mu^2-1\big)\big(\mu^2-a^2\big).
\end{aligned}
\eeQ
The Kerr-anti-de Sitter solution allows $| \mu |> |\kappa | ^{-1}$ and the metric has signature ($-1,1,1,1$) if
$\Delta_{\mu} >0$. If $m=0$, it has constant curvature $-\kappa^2$ and reduces to the anti-de Sitter spacetime.

In the region $-|\kappa | ^{-1} <\mu < |\kappa | ^{-1}$, $\lambda >0$, we can take the coordinate transformation
\beQ
\lambda=\hat{r}, \ \  \mu=a\cos\hat{\theta}, \ \ \chi=t-a\hat{\varphi}, \ \ \psi=\frac{1}{a}\hat{\varphi}
\eeQ
with $|\mu a|<1$ and it yields Boyer-Lindquist coordinates for the Kerr-anti-de Sitter spacetime
\beQ
\begin{aligned}
\widetilde{g}_{KAdS}=&-\Big[1-\frac{2m\hat{r}}{U} +\kappa^2\big(\hat{r}^2+a^2\sin^2\hat{\theta}\big)\Big]d\hat{t}^2
+\frac{U}{\Delta_{\hat{r}}}d\hat{r}^2 +\frac{U}{\Delta_{\hat{\theta}}}d\hat{\theta}^2\\
&+\frac{V}{U} \sin^2\hat{\theta} d\hat{\varphi}^2
-2a\sin^2\hat{\theta} \Big[\frac{2m\hat{r}}{U}-\kappa^2\big(\hat{r}^2+a^2\big)\Big]d\hat{t}d\hat{\varphi},
\end{aligned}
\eeQ
where
\beQ
\begin{aligned}
\Delta_{\hat{r}}&=\big(\hat{r}^2+a^2\big)\big(1+\kappa^2\hat{r}^2\big)-2m\hat{r},\quad \Delta_{\hat {\theta}}=1-\kappa^2a^2\cos^2\hat{\theta},\\
U&=\hat{r}^2+a^2\cos^2\hat{\theta},\quad V=2m\hat{r}a^2 \sin^2\hat{\theta} +U \big(\hat{r}^2+a^2\big)\big(1-\kappa^2a^2\big).
\end{aligned}
\eeQ
By \cite{HT}, we can know the total energy-momentum of $t$-slices
\beQ
E_0=\frac{m}{(1- \kappa^2 a^2)^2},\,\,
c_i= c'_i=0,\,\, J_1=J_2=0, \,\, J_3=\frac{m \kappa a}{(1- \kappa^2 a^2)^2}.
\eeQ

\mysection{Energy-momentum endomorphism }\ls

In this section we define energy-momentum endomorphisms for asymptotically anti-de Sitter spacetimes. Recall that the spinor bundle of the anti-de Sitter
spacetime is trivial and is $\hC ^4$ over the anti-de Sitter spacetime. The anti-de Sitter spacetime is characterized by imaginary Killing spinors satisfying
the following equations
\beQ
\widetilde{\nabla}^{AdS}_X \Phi_0+\frac{\kappa \sqrt{-1}}{2}X\cdot\Phi_0=0.
\eeQ
Denote $\hK$ the space of imaginary Killing spinors over the anti-de Sitter spacetime. It is a complex linear space with complex dimension $4$. There exists a one-to-one complex linear map
\beQ
\K : \hC ^4 \longrightarrow \hK.
\eeQ
For any given complex vector $\vec{\lambda}$, $\K (\vec{\lambda})=\Phi _0 ^{\lambda}$ is the unique corresponding Killing spinor.

We first define globally the energy-momentum endomorphism ${\bf Q}$ as a Hermitian transformation over complex space $\hC ^4$.
Let $\{\breve{e} _\alpha \}$ and $\breve{\nabla}$ be the frame and Levi-Civita connection of anti-de Sitter metric (\ref{ads-metric}) respectively.
For each end of an asymptotically anti-de Sitter initial data set,
\beQ
\begin{aligned}
\Theta = &\big(\breve{\nabla}^j g_{1j}-\breve{{\nabla}}_1tr_{\breve{g}}(g)\big)\mbox{Id} +
          \kappa \sum _{l}(a_{l1}-g_{l1}tr_{\breve{g}}(a))\sqrt{-1}\breve{e}_l             \\
         & -2\sum _l (h_{l1}-g_{l1}tr_{\breve{g}}(h))\breve{e}_0\cdot\breve{e}_l
\end{aligned}
\eeQ
serves as an endomorphism of the spinor bundle.

\begin{defn}
The energy-momentum endomorphism ${\bf Q}$ of an end for an asymptotically anti-de Sitter initial data set is a complex linear map
\beQ
{\bf Q}: \hC ^4 \longrightarrow \hC ^4
\eeQ
such that for any vector $\vec{\lambda} \in \hC ^4$,
\beQ
\langle \vec{\lambda}, {\bf Q}(\vec{\lambda})\rangle _{C} =\frac{1}{32\pi}\int _{S_\infty}\langle \Phi _0 ^\lambda, \Theta \cdot \Phi _0 ^\lambda \rangle\breve{\omega}
\eeQ
where $\langle \, , \rangle _{C}$ is the Hermitian inner product on $\hC ^4$, and $S_\infty$ is the 2-sphere at spatial infinity in $M$
and $\breve{\omega}$ is the reduced area form of $S_\infty$.
\end{defn}

Since $\Theta $ is Hermitian, ${\bf Q}$ is also Hermitian. Now we compute {\bf Q} explicitly under the following
Clifford representation. (We fix it for convenience throughout the paper although the whole results do not depend on the specific representation.)
\beq
\begin{aligned}
\breve{e}_0 \mapsto \begin{pmatrix}\ &\ &1 &\ \\ \ &\ & \ &1\\1&\ &\ &\ \\ \ &1&\ &\ \end{pmatrix}, &\quad
\breve{e}_1 \mapsto \begin{pmatrix}\ &\ &-1 &\ \\ \ &\ & \ &1\\1&\ &\ &\ \\ \ &-1&\ &\ \end{pmatrix},\\
\breve{e}_2 \mapsto \begin{pmatrix}\ &\ &\ &1 \\ \ &\ & 1 &\ \\\ &-1 &\ &\ \\ -1 &\ &\ &\ \end{pmatrix}, &\quad
\breve{e}_3 \mapsto \sqrt{-1}\begin{pmatrix}\ &\ &\ &1 \\ \ &\ & -1 &\ \\\ &-1 &\ &\ \\ 1 &\ &\ &\ \end{pmatrix}.
\label{repre}
\end{aligned}
\eeq
Under this representation, the imaginary Killing spinor $\Phi_0 ^\lambda$ is of the form
\begin{equation}\label{ik}
\Phi_0 ^\lambda =\begin{pmatrix}u^+e^{\frac{\kappa r}{2}}+u^-e^{-\frac{\kappa r}{2}}\\v^+e^{\frac{\kappa
r}{2}}+v^-e^{-\frac{\kappa r}{2}}\\-\sqrt{-1}u^+e^{\frac{\kappa r}{2}}+\sqrt{-1}u^-e^{-\frac{\kappa r}{2}} \\
\sqrt{-1}v^+e^{\frac{\kappa r}{2}}-\sqrt{-1}v^-e^{-\frac{\kappa r}{2}}
\end{pmatrix},
\end{equation}
where
\begin{eqnarray}\label{ik2}
\begin{aligned}
u^+ =&\Big(\lambda_1\cos\frac{\kappa t}{2}+\lambda_3\sin\frac{\kappa t}{2}\Big)
       e^{\frac{\sqrt{-1}}{2}\psi}\sin\frac{\theta}{2}\\
     &+\Big(\lambda_2\cos\frac{\kappa t}{2}+\lambda_4\sin\frac{\kappa t}{2}\Big)
       e^{\frac{-\sqrt{-1}}{2}\psi}\cos\frac{\theta}{2},\\
u^-=&\Big(-\lambda_1\sin\frac{\kappa t}{2}+\lambda_3\cos\frac{\kappa t}{2}\Big)
       e^{\frac{\sqrt{-1}}{2}\psi}\sin\frac{\theta}{2}\\
    &+\Big(-\lambda_2\sin\frac{\kappa t}{2}+\lambda_4\cos\frac{\kappa t}{2}\Big)
       e^{\frac{-\sqrt{-1}}{2}\psi}\cos\frac{\theta}{2},\\
v^+=&-\Big(-\lambda_1\sin\frac{\kappa t}{2}+\lambda_3\cos\frac{\kappa t}{2}\Big)
       e^{\frac{\sqrt{-1}}{2}\psi}\cos\frac{\theta}{2}\\
    &+\Big(-\lambda_2\sin\frac{\kappa t}{2}+\lambda_4\cos\frac{\kappa t}{2}\Big)
       e^{\frac{-\sqrt{-1}}{2}\psi}\sin\frac{\theta}{2}, \\
v^-=&-\Big(\lambda_1\cos\frac{\kappa t}{2}+\lambda_3\sin\frac{\kappa t}{2}\Big)
       e^{\frac{\sqrt{-1}}{2}\psi}\cos\frac{\theta}{2}\\
    &+\Big(\lambda_2\cos\frac{\kappa t}{2}+\lambda_4\sin\frac{\kappa t}{2}\Big)
       e^{\frac{-\sqrt{-1}}{2}\psi}\sin\frac{\theta}{2},
\end{aligned}
\end{eqnarray}
and $\lambda_1$, $\lambda_2$, $\lambda_3$ and $\lambda_4$ are four arbitrary complex numbers.

\begin{prop}
Under the Clifford multiplication (\ref{repre}), the energy-momentum endomorphism has the following form
\beq \label{Q}
\begin{aligned}
 {\bf Q} &=\begin{pmatrix}
 P          &     W \\
 \overline{W}^t  &    \hat{P}
 \end{pmatrix}, \qquad
 P=\begin{pmatrix}
 E_0-c_3             &      c_1-\sqrt{-1}c_2\\
 c_1+\sqrt{-1}c_2    &      E_0+c_3
 \end{pmatrix},\\
 W&=\begin{pmatrix}
 w_1  &     w_2 ^+\\
 w_2 ^-  &    -w_1
 \end{pmatrix}, \quad
 \hat{P}=\begin{pmatrix}
 E_0+c_3             &      -c_1+\sqrt{-1}c_2\\
 -c_1-\sqrt{-1}c_2   &      E_0-c_3
 \end{pmatrix},
\end{aligned}
\eeq
where $w_1 = c'_{3}-\sqrt{-1}J_{3}$, $w_2 ^\pm = -c'_{1} \pm J_{2}\pm \sqrt{-1}(c_{2}'\pm J_{1})$.
\end{prop}
\pf By (\ref{ik}), (\ref{ik2}), we have
\beQ
\begin{aligned}
\frac{1}{4}\int _{S_\infty}\langle \Phi _0 ^\lambda, \Theta \cdot \Phi _0 ^\lambda \rangle\breve{\omega}
=&\frac{1}{2}\int_{S_\infty}\mathcal{E}_1
            \big(\overline{u^+}u^++\overline{v^+}v^+\big) e^{\kappa r}\breve{\omega}\\
&+\int_{S_\infty}\mathcal{P}_{21}
            \big(\overline{u^+}v^+ +\overline{v^+}u^+\big)e^{\kappa r} \breve{\omega}\\
&+\sqrt{-1}\int_{S_\infty}\mathcal{P}_{31}
            \big(\overline{u^+}v^+ -\overline{v^+}u^+\big)e^{\kappa r}\breve{\omega},
\end{aligned}
\eeQ
and
\begin{equation*}
\begin{aligned}
\overline{u^+}u^+  +\overline{v^+}   v^+
=&\frac{1}{2}(\bar\lambda_1\lambda_1+\bar\lambda_2\lambda_2+\bar\lambda_3\lambda_3+\bar\lambda_4\lambda_4)\\
&+\frac{1}{2}\cos(\kappa t)\sin\theta \cos\psi(\bar\lambda_1\lambda_2+\bar\lambda_2\lambda_1-\bar\lambda_3\lambda_4-\bar\lambda_4\lambda_3)\\
&+\frac{1}{2}\sin(\kappa t)\sin\theta \cos\psi(\bar\lambda_1\lambda_4+\bar\lambda_2\lambda_3+\bar\lambda_3\lambda_2+\bar\lambda_4\lambda_1)\\
&+\frac{\sqrt{-1}}{2}\cos(\kappa t)\sin\theta \sin\psi(-\bar\lambda_1\lambda_2+\bar\lambda_2\lambda_1+\bar\lambda_3\lambda_4-\bar\lambda_4\lambda_3)\\
&+\frac{\sqrt{-1}}{2}\sin(\kappa t)\sin\theta \sin\psi(-\bar\lambda_1\lambda_4+\bar\lambda_2\lambda_3-\bar\lambda_3\lambda_2+\bar\lambda_4\lambda_1)\\
&+\frac{1}{2}\cos(\kappa t)\cos\theta(-\bar\lambda_1\lambda_1+\bar\lambda_2\lambda_2+\bar\lambda_3\lambda_3-\bar\lambda_4\lambda_4)\\
&+\frac{1}{2}\sin(\kappa t)\cos\theta(-\bar\lambda_1\lambda_3+\bar\lambda_2\lambda_4-\bar\lambda_3\lambda_1+\bar\lambda_4\lambda_2),
\end{aligned}
 \end{equation*}
\begin{equation*}
\begin{aligned}
\overline{u^+}v^+  +\overline{v^+}u^+
=&\frac{1}{2}\sin\theta(-\bar\lambda_1\lambda_3+\bar\lambda_2\lambda_4-\bar\lambda_3\lambda_1+\bar\lambda_4\lambda_2)\\
&+\frac{1}{2}\cos\psi(\bar\lambda_1\lambda_4-\bar\lambda_2\lambda_3-\bar\lambda_3\lambda_2+\bar\lambda_4\lambda_1)\\
&+\frac{\sqrt{-1}}{2}\sin\psi(-\bar\lambda_1\lambda_4-\bar\lambda_2\lambda_3+\bar\lambda_3\lambda_2+\bar\lambda_4\lambda_1)\\
&+\frac{1}{2}\cos(\kappa t)\cos\theta \cos\psi(-\bar\lambda_1\lambda_4-\bar\lambda_2\lambda_3-\bar\lambda_3\lambda_2-\bar\lambda_4\lambda_1)\\
&+\frac{1}{2}\sin(\kappa t) \cos\theta \cos\psi(\bar\lambda_1\lambda_2+\bar\lambda_2\lambda_1-\bar\lambda_3\lambda_4-\bar\lambda_4\lambda_3)\\
&+\frac{\sqrt{-1}}{2}\cos(\kappa t)\cos\theta \sin\psi(\bar\lambda_1\lambda_4-\bar\lambda_2\lambda_3+\bar\lambda_3\lambda_2-\bar\lambda_4\lambda_1)\\
&+\frac{\sqrt{-1}}{2}\sin(\kappa t)\cos\theta \sin\psi(-\bar\lambda_1\lambda_2+\bar\lambda_2\lambda_1+\bar\lambda_3\lambda_4-\bar\lambda_4\lambda_3),
\end{aligned}
 \end{equation*}
\begin{equation*}
\begin{aligned}
\overline{u^+}v^+ -\overline{v^+}u^+
=&\frac{1}{2}\sin\theta(-\bar\lambda_1\lambda_3+\bar\lambda_2\lambda_4+\bar\lambda_3\lambda_1-\bar\lambda_4\lambda_2)\\
&+\frac{1}{2}\cos(\kappa t)\cos\psi(\bar\lambda_1\lambda_4-\bar\lambda_2\lambda_3+\bar\lambda_3\lambda_2-\bar\lambda_4\lambda_1)\\
&+\frac{1}{2}\sin(\kappa t)\cos\psi(-\bar\lambda_1\lambda_2+\bar\lambda_2\lambda_1+\bar\lambda_3\lambda_4-\bar\lambda_4\lambda_3)\\
&+\frac{\sqrt{-1}}{2}\cos(\kappa t)\sin\psi(-\bar\lambda_1\lambda_4-\bar\lambda_2\lambda_3-\bar\lambda_3\lambda_2-\bar\lambda_4\lambda_1)\\
&+\frac{\sqrt{-1}}{2}\sin(\kappa t)\sin\psi(\bar\lambda_1\lambda_2+\bar\lambda_2\lambda_1-\bar\lambda_3\lambda_4-\bar\lambda_4\lambda_3)\\
&+\frac{1}{2}\cos\theta \cos\psi(-\bar\lambda_1\lambda_4-\bar\lambda_2\lambda_3+\bar\lambda_3\lambda_2+\bar\lambda_4\lambda_1)\\
&+\frac{\sqrt{-1}}{2}\cos\theta \sin\psi(\bar\lambda_1\lambda_4-\bar\lambda_2\lambda_3-\bar\lambda_3\lambda_2+\bar\lambda_4\lambda_1).
\end{aligned}
 \end{equation*}
Thus we obtain
 \beQ
\langle \vec{\lambda}, {\bf Q}(\vec{\lambda})\rangle _{C } =\big(\bar\lambda_1, \bar\lambda_2, \bar\lambda_3,\bar\lambda_4\big)
      {\bf Q}\big(\lambda_1, \lambda_2, \lambda_3, \lambda_4 \big)^t,
\eeQ
where ${\bf Q}$ is given by (\ref{Q}). \qed

Denote ${\bf c} =(c_1, c_2, c_3)$, ${\bf c}' =(c' _1, c' _2, c' _3)$, ${\bf J} =(J_1, J_2, J_3)=\vec{j}$ and
\beq\label{3constants}
\begin{aligned}
L=&\big(|{\bf c}|^2 +|{\bf c}'|^2 +|{\bf J}|^2 \big)^{\frac{1}{2}},\\
A=&\big(|{\bf c} \times {\bf c}' |^2+|{\bf c} \times {\bf J}|^2 +| {\bf c}'\times {\bf J} |^2\big)^{\frac{1}{4}},\\
V=&\big(\varepsilon_{ijl}c_i c_{j}' J_l \big) ^{\frac{1}{3}},
\end{aligned}
\eeq
where $2L$, $2A^2$ and $V^3$ are the (normalized) length, surface area and volume of the parallelepiped spanned by
${\bf c}$, ${\bf c}'$ and ${\bf J}$. Clearly, $L^2 \geq 3V^2$. Using (\ref{relations}), we can prove
\beQ
\begin{aligned}
L=&\big(|\vec{m}|^2 +|{\vec c}|^2 +|{\vec j}|^2 \big)^{\frac{1}{2}},\\
A=&\big(|\vec{c} \times \vec{m}|^2+|\vec{c} \times \vec{j}|^2 +|\vec{m}\times \vec{j} |^2 \big)^{\frac{1}{4}},\\
V=&\big(-\varepsilon_{ijl}m_{(i)} c_{(j)} j_{(l)} \big) ^{\frac{1}{3}}.
\end{aligned}
\eeQ

Note that $\R^{3,2}$ has two timelike Killing vectors $\frac{\partial}{\partial y^0}$,  $\frac{\partial}{\partial y^4}$
and three spacelike Killing vectors $\frac{\partial}{\partial y^1}$, $\frac{\partial}{\partial y^2}$, $\frac{\partial}{\partial y^3}$.
Physically, $E$ measures the rotation on the plane $(y^0, y^4)$, $c_i$ measures the rotation on the plane $(y^i, y^4)$,
$c' _i$ measures the rotation on the plane $(y^0, y^i)$ and $J_i$ measures the rotation on the plane $(y^j, y^l)$ where $\{i,j,k\}$ is the even
permutation of $\{1,2,3\}$. But these rotations are all observed from a curved space, the hyperboloid (\ref{ads}), so they contain both
translation and rotation of an asymptotically anti-de Sitter spacetime. This indicates that we can not simply refer them as the center of mass
as well as the total angular momentum. The total effect of translation and rotation is given by the parallelepiped spanned by
${\bf c}$, ${\bf c}'$ and ${\bf J}$ which can be measured from its length of the edges, surface area and the volume.

Denote by $tr{\bf Q}$, ${\bf Q} ^{(2)}$, ${\bf Q} ^{(3)}$ and $\det{\bf Q}$ the trace, sum of the second-order minors,
sum of the third-order minors and the determinant of ${\bf Q}$. It is straightforward to prove the following proposition.
\begin{prop}\label{detQ}
\beQ
\begin{aligned}
tr{\bf Q}=&4E_0, \quad {\bf Q} ^{(2)}=6E_0 ^2 -2L^2, \\
{\bf Q} ^{(3)}=&4E_0 (E_0^2-L^2)+8V^3,\\
\det {\bf Q} =&\big(E_0^2-L^2 \big)^2+8 E_0 V^3-4A^4,
\end{aligned}
\eeQ
and they are independent on $t$. Moreover, they are independent on specific Clifford representation also.
\end{prop}

\mysection{The positive energy theorem}\ls

Now we prove the positive energy theorem for Henneaux-Teitelboim's total energy-momentum.
Let $(M,g,h)$ be an asymptotically anti-de Sitter initial data set in $(N,\widetilde{g})$ which satisfies the dominant
energy condition (\ref{dec}). Let $\nabla$ and $\widetilde{\nabla}$
be the Levi-Civita connections of $g$ and $\widetilde{g}$ respectively. Let $\mathbb{S}$ be the locally
spinor bundle of $N$ and we still denote by $\mathbb{S}$ its restriction to $M$. Since the hypersurface $M$ is three
dimensional, the restriction $\mathbb{S}$ is globally defined on $M$. And we lift $\nabla$ and $\widetilde{\nabla}$
to $\mathbb{S}$ and denote the corresponding spin connections the same as $\nabla$ and $\widetilde{\nabla}$.
Fix a point $p\in M$ and an orthonormal basis $\{e_\alpha\}$ of $T_pN$ with
$e_0$ normal and $\{e_i\}$ tangent to $M$. Extend $\{e_\alpha\}$ to a local orthonormal frame in a neighborhood of
$p$ in $M$ such that $(\nabla^g_ie_j)_p=0$. Extend this to a local orthonormal frame $\{e_\alpha\}$ for $N$ with
$(\widetilde{\nabla}_0e_j)_p=0$. Then $(\widetilde{\nabla}_ie_j)_p=h_{ij}e_0$, $(\widetilde{\nabla}_ie_0)_p=h_{ij}e_j$.
Define
\beQ
\widehat{\nabla}_i=\widetilde{\nabla}_i+\frac{\sqrt{-1}}{2}\kappa e_i, \quad \widehat D =\sum _{i=1} ^3 e_i \widehat{\nabla}_i.
\eeQ
Recall that the Weitzenb\"{o}ck formula gives (e.g. \cite{XZ})
\begin{eqnarray}\label{WI}
\begin{aligned}
\int_M|\widehat\nabla\phi|^2 -|\widehat D\phi|^2+ \langle\phi,\widehat{\mathcal{R}}\phi\rangle
=\int_{\partial M}\langle\phi,\sum_{ j\neq i }e_i\cdot e_j\cdot\widehat\nabla_j \phi\rangle\ast e^i
\end{aligned}
\end{eqnarray}
where $\widehat{\mathcal{R}}=\frac{1}{2}(T_{00}-T_{0i}e_0 e_i)$ and $\langle\cdot, \cdot\rangle$ is the positive definite
inner product on the spinor bundle $\mathbb{S}$ under which $e_0 \cdot$ is Hermitian and $e_i \cdot$ is skew-Hermitian.

Now we briefly review some basic facts in  \cite{Wi, Wa, CH, Z, M, CMT}. Note that $g=\breve g + a$ with
$a=O(e^{-\tau\kappa r})$, $\breve \nabla a=O(e^{-\tau\kappa r})$, and $\breve \nabla \breve \nabla a=O(e^{-\tau\kappa r})$.
Orthonormalizing $\breve e_i$ gives a gauge transformation $$\mathcal{A}: SO(\breve g) \rightarrow SO(g)$$
$$\breve e_i \mapsto e_i$$ (and in addition $\breve{e}_0 \mapsto \breve{e}_0$) which identifies the corresponding spin group and
the spinor bundles. Moreover,
\beQ
e_i=\breve e_i-\frac{1}{2}a_{ik}\breve e_k+o(e^{-\tau\kappa r}).
\eeQ

We extend the imaginary Killing spinors $\Phi_0$ (\ref{ik}) on the
end to the inside smoothly. With respect to the metric $g$, these
imaginary Killing spinors $\Phi_0$ can be written as
$\overline{\Phi}_0=\mathcal{A}\Phi_0$.

We try to find the unique solution $\widehat D\phi=0$ such that $\phi$
is asymptotic to the imaginary Killing spinors $\overline{\Phi}_0$ on certain end, and to zero on the other ends.
Let $C_0^\infty(\mathbb{S})$ be the space of smooth sections of
the spinor bundle $ \mathbb{S}$ with compact support.
Let the Hilbert space $H^1(\mathbb{S})$ be the closure of $C_0^\infty(\mathbb{S})$
with respect to the $W^{1,2}$ inner product. Now the bounded
bilinear form $\mathcal{B}$ defined on $C_0^\infty(\mathbb{S})$ satisfies
\beQ
\mathcal{B}(\phi,\psi):=\int_M \langle\widehat D\phi, \widehat D\psi \rangle=\int_M |\widehat\nabla\phi|^2 +\langle \phi,\widehat{\mathcal{R}}\phi\rangle
\geq \int_M |\widehat\nabla\phi|^2
\eeQ
by the Weitzenb\"{o}ck formula (\ref{WI}) and the dominant energy condition (\ref{dec}). Thus we can extend $\mathcal{B}(\cdot,\cdot)$ to $H^1(\mathbb{S})$ as a coercive bilinear form.
This is a consequence of the Poincar\'{e} inequality.

\begin{lem}\label{Dirac}
Let $(M,g,h)$ be a 3-dimensional asymptotically anti-de Sitter initial data set in spacetime $(N,\widetilde{g})$. Suppose $(N,\widetilde{g})$ satisfies the dominant energy condition.
Then there exists a unique spinor $\Phi_1$ in $H^1(\mathbb{S})$  such that
\beQ
\widehat D(\Phi_1+\overline{\Phi}_0)=0.
\eeQ
\end{lem}
\pf The proof is essentially similar to that of Lemma 5.1 in \cite{XZ}. Since
$\mathcal{B}(\cdot,\cdot)$ is coercive on $H^1(\mathbb{S})$, and
$\widehat D \overline{\Phi}_0 \in L^2(\mathbb{S})$, $\widehat\nabla \overline{\Phi}_0 \in L^2(\mathbb{S})$.
By the theorem of Lax-Milgram, there exists a spinor $\Phi_1 \in H^1(\mathbb{S})$ such that
$\widehat D^\ast\widehat D\Phi_1=-\widehat D^\ast \widehat D\overline{\Phi}_0$
weakly. Here $\widehat D^\ast$ is the formal adjoint operator of $\widehat D$. Let $\phi=\Phi_1+\overline{\Phi}_0$ and
$\psi=\widehat D\phi$. The elliptic regularity tells us
that $\psi \in H^1(\mathbb{S})$, and
$\widehat D^\ast\psi=0$
in the classical sense \cite{BC}. The Weitzenb\"{o}ck formula implies that
$\widehat \nabla \psi=0$. We thus have $|\partial_i \log
|\psi|^2| \leq \kappa+|h|$ on the complement of the zero set of
$\psi$ on $M$. If there exists $x_0 \in M$ such that
$|\psi(x_0)|\neq 0$, then integrating it along a path from $x_0\in
M$ gives
$$|\psi(x)|^2 \geq |\psi(x_0)|^2e^{(\kappa+|h|)(|x_0|-|x|)}.$$
Obviously, $\psi$ is not in $L^2(\mathbb{S})$ which gives the
contradiction. Hence $\psi=0$, and the proof of this lemma is
complete.\qed

Now let $\phi$ be the solution of the Dirac-type equation
$\widehat D\phi=0$ as in Lemma \ref{Dirac}. Plugging this
$\phi$ into the Weitzenb\"{o}ck formula (\ref{WI}), we obtain that
the boundary term is nonnegative under the dominant energy condition (\ref{dec}).
Using the Clifford representation (\ref{repre}) and (\ref{ik}) for $\Phi_0$, in the polar coordinates,
the boundary term of the Weitzenb\"{o}ck formula (\ref{WI}) in the right hand side gives
\begin{eqnarray*}
\begin{aligned}
RHS\,(\ref{WI})
          =&\frac{1}{4}\lim_{r\rightarrow \infty}\int_{S_r}
          (\breve{\nabla}^j g_{1j}-\breve{{\nabla}}_1tr_{\breve{g}}(g))|\Phi_0|^2\breve{\omega}\\
          &+\frac{1}{4}\lim_{r\rightarrow \infty}\int_{S_r}
          \kappa(a_{k1}-g_{k1}tr_{\breve{g}}(a))\langle\Phi_0,\sqrt{-1}\breve{e}_k\cdot\Phi_0 \rangle\breve{\omega}\\
          &-\frac{1}{2}\lim_{r\rightarrow\infty}\int_{S_r}
          (h_{k1}-g_{k1}tr_{\breve{g}}(h))\langle\Phi_0,\breve{e}_0\cdot\breve{e}_k\cdot\Phi_0\rangle\breve{\omega}\\
          =&8 \pi\langle \vec{\lambda}, {\bf Q}(\vec{\lambda})\rangle _{C}.
          \end{aligned}
\end{eqnarray*}

Now we prove our main theorem.
\begin{thm}\label{pmt}
Let $(M,g,h)$ be a 3-dimensional asymptotically anti-de Sitter initial data set in spacetime $(N,\widetilde{g})$. Suppose $(N,\widetilde{g})$ satisfies the dominant energy condition. Then, for each end
 \beq\label{a}
E_0 \geq \sqrt{L^2 -2V^2 +2 \big(\max\{A^4 -L^2 V^2, 0\}\big)^\frac{1}{2} }. \label{WXZ-em}
 \eeq
If $E_0=0$ for some end, then $M$ has only one end, ${\bf Q}=0$, and $(N,\widetilde{g})$ is anti-de Sitter along $M$.
\end{thm}
\pf Let $\phi$ be the solution of the Dirac-type equation $\widehat D\phi=0$ as in Lemma \ref{Dirac}. The dominant energy condition (\ref{dec}) ensures that ${\bf Q}$ is positive semidefinite. Now the trace yields
\beQ
E _0 \geq 0.
\eeQ
The sum of the second-order principal minors yields
\beQ
E _0 ^2 \geq L ^2 /3.
\eeQ
Therefore
\beQ
V^3 \leq V^2 L/\sqrt{3} \leq V^2 E_0,
\eeQ
The sum of the third-order principal minors yields
\beQ
0 \leq E_0 \big(E_0^2-L^2\big)+2V^3.
\eeQ
So, if $E_0 >0$, it implies
\beQ
E_0 ^2  \geq L^2 -2V^2\geq L^2 -2L^2 /3=L^2/3.
\eeQ
Now we use the nonnegativity of the determinant of ${\bf Q}$ to prove (\ref{a}). Since
\beQ
2 E_0 V^3 \leq \big(E _0 ^2 + V^2\big) V^2,
\eeQ
we obtain
\beQ
0\leq \det {\bf Q} \leq \big(E_0^2-L^2 +2 V^2 \big)^2 -4(A^4 -L^2 V^2).
\eeQ
This implies (\ref{a}).

If $E_0=0$ for some end, then it is straightforward that $M$ has only one end, and ${\bf Q}=0$. This implies that there exists $\{\phi_\alpha\}$ which forms a basis of the spinor bundle everywhere over $M$ such that $\widehat{\nabla}\phi_\alpha=0$. Standard argument gives
\beQ
\widetilde {R}_{ijkl}=(-\kappa^2)\big(\widetilde {g}_{ik}\widetilde {g}_{jl}-\widetilde {g}_{il}\widetilde {g}_{jk}\big),\qquad \widetilde {R}_{0jkl}=0
\eeQ
along $M$. The Einstein field equations (\ref{EinsteinEqs}) yield
\beQ
T_{00}=\widetilde {R}_{00}+\frac{1}{2}\widetilde{R}-\Lambda=\frac{1}{2} \sum_{i,j}R_{ijij}-\Lambda=0.
\eeQ
Then (\ref{dec}) implies $T_{\alpha\beta}=0$ and furthermore
\beQ
\widetilde {R}_{0j0l}=\kappa^2\widetilde {g}_{jl}.
\eeQ
Therefore, the curvature tensors of $(N,\widetilde{g})$ are
\beQ
\widetilde{R}_{\alpha\beta\gamma\delta}=(-\kappa^2)\big(\widetilde {g}_{\alpha\gamma}\widetilde {g}_{\beta\delta}-\widetilde
{g}_{\alpha\delta}\widetilde {g}_{\beta\gamma}\big)
\eeQ
and $N$ is anti-de Sitter along $M$. \qed

\begin{coro}
If three vectors ${\bf c}$, ${\bf c}'$, ${\bf J}$ or $\vec{m}$, $\vec{c}$, $\vec{j}$ are linearly dependent,
i.e, $V=0$, then the energy-momentum inequality (\ref{WXZ-em}) becomes
\beQ
E_0 \geq \sqrt{L^2 +2 A^2}.
\eeQ
\end{coro}

This corollary generalizes the energy-momentum inequality (\ref{i}). It also indicates that $ \vec{m}$, $\vec{c}$ and $\vec{j}$ play the same role in physics.

\begin{rmk}
In the above energy-momentum endomorphism ${\bf Q}$, nonnegativity of the second-order minor $K$ gives $E_0 \geq |{\bf c}|$. However, this inequality does depend on the Clifford representation. For instance, if we permute $\breve{e}_1 \rightarrow \breve{e}_2$, $\breve{e}_2 \rightarrow \breve{e}_3$, $\breve{e}_3 \rightarrow \breve{e}_1$ in Clifford representation (\ref{repre}), the energy-momentum endomorphism ${\bf Q}$ will change to the new one with
\beQ
\begin{aligned}
 P=&\begin{pmatrix}
 E_0+c_3 -c'_1-J_2             &      -c' _2+J_1\\
 -c' _2+J_1   &      E_0+c_3+c'_1+J_2
 \end{pmatrix},\\
 W=&\begin{pmatrix}
 c_1 +c' _3  &     c_2 + J_3\\
 c_2 - J_3 &    -c_1 +c' _3
 \end{pmatrix}, \\
 \hat{P}=&\begin{pmatrix}
  E_0-c_3 +c'_1-J_2             &      c'_2+J_1\\
 c'_2+J_1   &      E_0-c_3-c'_1+J_2
 \end{pmatrix}.
\end{aligned}
\eeQ
The inequality $E_0 \geq |{\bf c}|$ does not hold in the new energy-momentum endomorphism.
\end{rmk}

If $V>0$, $E_0$ is very close to $V$ and $|\vec{c}|$, $|\vec{j}|$ are sufficiently small, the universal inequality
$L^2 \geq 3V^2$ will give $|\vec{m}| > m_{0}$.
\begin{rmk}\label{positive-energy4}
When three vectors $\vec{m}$, $\vec{c}$, $\vec{j}$ are linearly independent, that ${\bf Q}$ is positive semidefinite, in genral,
does not result that the total energy four vector $m_{(\mu)}$ is timelike.
\end{rmk}

\begin{rmk}
One can construct certain regular initial initial data sets which are Kerr-anti-de Sitter at infinity with
\begin{eqnarray*}
E_0 =1, \quad {\bf c} =(0, 0, 0), \quad {\bf c}' =(0, 0, 0), \quad {\bf J} =(0, 0, \kappa a)
\end{eqnarray*}
where $1>\kappa |a|$. Thus (\ref{WXZ-em}) is optimal in this sense.
\end{rmk}

If $M$ has a future/past trapped surface $(\Sigma,\bar{g},\bar{h})$ equipped with the induced metric $\bar{g}$ and the
second fundamental form $\bar{h}$
\beQ
tr_{\bar g} (\bar h) \mp tr_{\bar g} (h|_\Sigma ) \geq 0.
\eeQ
Let $e_3$ be outward normal and $e_A$ be tangent to $\Sigma$. The boundary term involving $\Sigma $ in the Weitzenb\"{o}ck formula is
\begin{eqnarray*}
\int_\Sigma \langle \phi, e_3 e_A \widehat\nabla_A \phi \rangle = \int_\Sigma \langle \phi, e_3 e_A \widetilde\nabla_A \phi \rangle
-\int_\Sigma \langle \phi, \sqrt{-1}\kappa e_3 \phi \rangle.
\end{eqnarray*}
\begin{rmk}
Theorem \ref{pmt} also holds for black holes. This is because that, under the local boundary conditions, the term $\langle\phi, e_3 \phi\rangle$
is both imaginary and real, hence zero. Then it follows by the standard argument \cite{GHHP}.
\end{rmk}

\mysection{Geometric invariant}

We shall show that the determinant of the total energy-momentum endomorphism ${\bf Q}$ is the geometric invariant which is independent on the choice
of admissible asymptotic coordinates.

We omit the upper-case $HT$ and denote $J_{ab}$ ($0 \leq a, b\leq 4$) as Henneaux-Teitelboim's total energy-momentum in this section. It yields
two $O(3,2)$ Casimir invariants \cite{HT}
\beQ
\begin{aligned}
I_1 &=\frac{1}{2}J_{ab}J^{ab}=-\frac{1}{2}J_{a}^{\ b}J_{b}^{\ a},\\
I_2 &=\frac{1}{2}J_{a}^{\ b}J_{b}^{\ c}J_{c}^{\ d}J_{d}^{\ a}-\frac{1}{4}(J_{a}^{\ b}J_{b}^{\ a})^2.
\end{aligned}
\eeQ

\begin{thm}
Denote $\det{\bf Q}$ as the determinant of the energy-momentum endomorphism ${\bf Q}$. We have
\beq
\det {\bf Q}=\Big(\frac{\kappa}{16\pi}\Big)^4 \big(I_1^2+2I_2 \big).  \label{inv-detQ}
\eeq
\end{thm}
\pf It is straightforward that
\begin{eqnarray*}
\begin{aligned}
\det {\bf Q}
 =&\big(E_0^2-L^2 \big)^2+8 E_0 V^3-4A^4\\
=&E_0^4+\sum_{i}\big[c_{i}^4+(c'_{i})^4+J_{i}^4\big]\\
&+\sum_{i\neq j}\big[c_{i}^2c_{j}^2+(c'_{i})^2(c'_{j})^2+J_{i}^2J_{j}^2-2c_{i}^2J_{j}^2-2(c'_{i})^2J_{j}^2\big]\\
&-2E_0^2\sum_{i}\big[c_{i}^2+(c'_{i})^2\big]+2\sum_{i}c_{i}^2(c'_{i})^2\\
&-2\sum_{i\neq j}c_{i}^2(c'_{j})^2-2E_0^2\sum_{i}J_{i}^2+2\sum_{i}\big[c_{i}^2J_{i}^2+(c'_{i})^2J_{i}^2\big]\\
&+8E_0 \varepsilon_{ijk}c_{i}c'_{j}J_k+4\sum_{i\neq j}\big(c_{i}c_{j}c'_{i}c'_{j}+c_{i}c_{j}J_{i}J_{j}+c'_{i}c'_{j}J_{i}J_{j} \big).
\end{aligned}
\end{eqnarray*}
By (\ref{em-and-HT}), we obtain that, in the right hand side of above equality, the sum of the first and the second terms is equal to I,
the sum of the third, the forth and the fifth terms is equal to II, the sum of the sixth, the seventh and the eighth terms is equal to III,
and the sum of the ninth and the tenth terms is equal to IV. Thus
\beQ
\Big(\frac{16\pi}{\kappa}\Big)^4\det {\bf Q}=\mbox{I}+\mbox{II}+\mbox{III}+\mbox{IV},
\eeQ
where
\beQ
\begin{aligned}
\mbox{I}=&\sum_{a< b}J_{ab}^4,\\
\mbox{II}=&\sum_{a,\ b,\ c \ \mbox{distinct}}J_{a}^{\ b}J_{b}^{\ a}J_{a}^{\ c}J_{c}^{\ a},\\
\mbox{III}=&-\frac{1}{4}\sum_{a,\ b,\ c,\ d \ \mbox{distinct}}J_{a}^{\ b}J_{b}^{\ a}J_{c}^{\ d}J_{d}^{\ c},\\
\mbox{IV}=&\sum_{a,\ b,\ c,\ d \ \mbox{distinct}}J_{a}^{\ b}J_{b}^{\ c}J_{c}^{\ d}J_{d}^{\ a}.
\end{aligned}
\eeQ

On the other hand,
\beQ
\begin{aligned}
J_{a}^{\ b}J_{b}^{\ c}J_{c}^{\ d}J_{d}^{\ a}=&2\mbox{I}+2\mbox{II}+\mbox{IV},\\
\frac{1}{4}J_{a}^{\ b}J_{b}^{\ a}J_{d}^{\ c}J_{c}^{\ d}=&\mbox{I}+\mbox{II}-\mbox{III}.
\end{aligned}
\eeQ
Therefore we obtain (\ref{inv-detQ}). \qed

Let $(t,r, \theta ^A)$, $(\hat{t},\hat{r}, \hat{\theta }^A)$ be two asymptotic coordinates of $M$ on the end,
where $\{\theta ^A \}=\{\theta, \psi\}$, $\{\hat{\theta} ^A \}=\{\hat\theta, \hat\psi\}$. We say that the coordinate transformation
\beQ
\D: (t,r, \theta ^A) \longrightarrow (\hat{t},\hat{r}, \hat{\theta }^A)
\eeQ
is {\em admissible} if, for $r$ sufficiently large, that
\beq \label{admissible}
\begin{aligned}
\hat t = t +o(e^{-\frac{5\kappa}{2}r}), \quad & \breve{e} _0 (\hat t)=\breve{e} _0 {(t)} +o(e^{-\frac{7\kappa}{2}r}),\\
\hat r = r +o(e^{-\frac{3\kappa}{2}r}), \quad & \breve{e} _1 (\hat r)=\breve{e} _1 {(r)} +o(e^{-\frac{3\kappa}{2}r}),\\
\hat \theta ^A = \theta ^A +o(e^{-\frac{5\kappa}{2}r}), \quad & \breve{e} _B (\hat \theta ^A)=\breve{e} _B {(\theta ^A)} +o(e^{-\frac{7\kappa}{2}r}).
\end{aligned}
\eeq

\begin{prop}
The admissible coordinate transformations on ends will preserve
Henneaux-Teitelboim's total energy-momentum .
\end{prop}
\pf The proof is essentially the same as that of Theorem 2.3 \cite{CN}, where it is used that $X=U_{\alpha \beta}$ is a Killing vector. So the proof goes through no matter that $t$ is zero or not.  \qed

\begin{thm}
The determinant $\det{\bf Q}$ of the energy-momentum endomorphism ${\bf Q}$ is invariant under admissible coordinate transformation (\ref{admissible}) on ends.
It serves as the geometric invariant of asymptotically anti-de Sitter spacetimes.
\end{thm}

\begin{rmk}
We may define $\sqrt[4]{\det{\bf Q}}$ as the total rest mass of asymptotically anti-de Sitter spacetimes.
\end{rmk}


{\footnotesize {\it Acknowledgement.} The authors are indebted to S. Deser for valuable comments.
This work is partially supported by the National Science Foundation of China (grants 11021091, 11171328, 11121101, 11401168)
and the Innovation Program of Shanghai Municipal Education Commission Grant 11ZZ01. The paper is part of the first author's PhD thesis
at the Institute of Mathematics, Academy of Mathematics and Systems Science, Chinese Academy of Sciences. He would like to thank the institute
for the hospitality and constant support.}


\mysection{Appendix A: Ten Killing vectors for AdS spacetime}
The followings are ten Killing vectors $U_{\alpha\beta}$ generating rotations for $\R^{3,2}$ along $t$-slices.
\beQ
\begin{aligned}
 U_{10}=&\frac{\cos (\kappa t)}{\kappa} \Big[\sin\theta \cos\psi\frac{\partial}{\partial r}
        +\kappa \coth(\kappa r)\Big(\cos\theta\cos\psi\frac{\partial}{\partial \theta}
        -\frac{\sin\psi}{\sin \theta}\frac{\partial}{\partial\psi}\Big)\Big]\\
        &-\frac{\sin (\kappa t)}{\kappa} \tanh(\kappa r)\sin\theta \cos\psi\ \frac{\partial}{\partial t},\\
 U_{20}=&\frac{\cos (\kappa t)}{\kappa} \Big[\sin\theta \sin\psi\frac{\partial}{\partial r}
        +\kappa \coth(\kappa r)\Big(\cos\theta\sin\psi\frac{\partial}{\partial \theta}
        +\frac{\cos\psi}{\sin \theta}\frac{\partial}{\partial\psi}\Big)\Big]\\
        &-\frac{\sin (\kappa t)}{\kappa}\tanh(\kappa r)\sin\theta \sin\psi \frac{\partial}{\partial t},\\
 U_{30}=&\frac{\cos (\kappa t)}{\kappa}\Big[\cos\theta\frac{\partial}{\partial r}
        -\kappa \coth(\kappa r)\sin\theta\frac{\partial}{\partial \theta}\Big]
        -\frac{\sin (\kappa t)}{\kappa}\tanh(\kappa r)\cos\theta\frac{\partial}{\partial t},\\
 U_{40}=&\frac{1}{\kappa} \frac{\partial}{\partial t},\\
 U_{14}=&\frac{\sin (\kappa t)}{\kappa}  \Big[\sin\theta \cos\psi\frac{\partial}{\partial r}
        +\kappa \coth(\kappa r)\Big(\cos\theta\cos\psi\frac{\partial}{\partial \theta}
        -\frac{\sin\psi}{\sin\theta}\frac{\partial}{\partial \psi}\Big)\Big]\\
        &+\frac{\cos (\kappa t)}{\kappa}\tanh(\kappa r) \sin\theta \cos\psi\frac{\partial}{\partial t},\\
 U_{24}=& \frac{\sin (\kappa t)}{\kappa} \Big[\sin\theta \sin\psi\frac{\partial}{\partial r}
        +\kappa \coth(\kappa r)\Big(\cos\theta\sin\psi\frac{\partial}{\partial \theta}
        +\frac{\cos\psi}{\sin\theta}\frac{\partial}{\partial \psi}\Big)\Big]\\
        &+\frac{\cos (\kappa t)}{\kappa}\tanh(\kappa r)\sin\theta \sin\psi\frac{\partial}{\partial t},\\
 U_{34}=&\frac{\sin (\kappa t)}{\kappa}\Big[\cos\theta\frac{\partial}{\partial r}
        -\kappa \coth(\kappa r)\sin\theta\frac{\partial}{\partial \theta}\Big]
        +\frac{\cos (\kappa t)}{\kappa}\tanh(\kappa r)\cos\theta\frac{\partial}{\partial t},\\
 U_{12}=&\frac{\partial}{\partial \psi},\\
 U_{23}=&-\sin\psi\frac{\partial}{\partial\theta}
         -\frac{\cos\theta \cos \psi}{\sin \theta}\frac{\partial}{\partial \psi},\\
 U_{31}=&\cos\psi\frac{\partial}{\partial \theta}
        -\frac{\cos\theta \sin\psi}{\sin \theta}\frac{\partial}{\partial \psi}.
\end{aligned}
\eeQ

\mysection{Appendix B: Center of AdS mass coordinates}

We shall explicitly construct a $SO(3,1)$ coordinate transformation on $t=0$ slice to change arbitrary coordinates $\{y ^\alpha\}$
to the center of AdS mass coordinates if the mass vector is timelike. Denote the $SO(3,1)$ matrix $\B=(B^{\alpha}_{\ \beta} )$. When $t=0$, the admissible coordinate transformations reduce to
\beQ
z^\alpha =B^{\alpha}_{\ \beta}y ^\beta, \qquad z^4= y^4.
\eeQ
Denote the lower-bar terms by the corresponding quantities in new coordinates. Since $z_\alpha=B_{\alpha}^{\ \beta}y_\beta$,
$B^{\alpha}_{\ \beta}=\eta^{\alpha\gamma}B_{\gamma}^{\ \delta} \eta_{\delta \beta}$ for the flat metric $\eta $ on $\R^{3,2}$, we have
\begin{eqnarray*}
\begin{aligned}
\underline{U} _{4\alpha}=&z_4\frac{\partial}{\partial z^\alpha}-z_\alpha\frac{\partial}{\partial z^4}
=B_{\alpha}^{\ \beta}U_{4\beta},\\
\underline{U}_{i0}=&z_i\frac{\partial}{\partial z^0}-z_0\frac{\partial}{\partial z^i}
=B_{0}^{\ \alpha}B_{i}^{\ \beta}U_{\beta \alpha},\\
\underline{U}_{ij}=&z_i\frac{\partial}{\partial z^j}-z_j\frac{\partial}{\partial z^i}
=B_{j}^{\ \alpha}B_{i}^{\ \beta}U_{\beta \alpha}.
\end{aligned}
\end{eqnarray*}
Consequently, the mass vector $m_{(\mu)}$, the center of mass $c_{(i)}$, and the angular momentum $J_{(i)(j)}$ defined in \cite{CMT}
have the following transformation laws
\begin{eqnarray}\label{relation}
\begin{aligned}
\underline{m}_{(\alpha)}=&B_{\alpha}^{\ \beta}m_{(\beta)},\\
\underline{c}_{(i)}=&(B_{i}^{\ j}B_{0}^{\ 0}-B_{i}^{\ 0}B_{0}^{\ j})c_{(j)}+B_{i}^{\ j}B_{0}^{\ k}J_{(j)(k)},\\
\underline{J}_{(i)(j)}=&(B_{i}^{\ k}B_{j}^{\ 0}-B_{i}^{\ 0}B_{j}^{\ k})c_{(k)}+B_{i}^{\ k}B_{j}^{\ l}J_{(k)(l)}.
\end{aligned}
\end{eqnarray}
By (\ref{relation}), we find that
the following $SO(3,1)$ matrix $\B_1$ changing the vector $(m_{(0)},m_{(1)}, m_{(2)}, m_{(3)})$ to $(\| m_\mu\|,0,0,0)$ if it is timelike,
\begin{equation*} \label{B1}
\B_1=
\begin{pmatrix}
\frac{m_{(0)}}{\| m_\mu\|}       &      -\frac{m_{(1)}}{\| m_\mu\|}        &      -\frac{m_{(2)}}{\| m_\mu\|}        &       -\frac{m_{(3)}}{\| m_\mu\|}  \\
-\frac{| m_i|}{\| m_\mu\|}         &      \frac{m_{(0)} m_{(1)}}{| m_i| \| m_\mu\|}            &      \frac{m_{(0)}m_{(2)}}{| m_i| \| m_\mu\|}                   &       \frac{m_0m_3}{| m_i| \| m_\mu\|}    \\
0               &       -\frac{m_{(2)}}{\sqrt{m_{(1)}^2+m_{(2)}^2}}                 &      \frac{m_{(1)}}{\sqrt{m_{(1)}^2+m_{(2)}^2}}          &       0   \\
0               &      -\frac{m_{(1)}m_{(3)}}{\sqrt{m_{(1)}^2+m_{(2)}^2}| m_i|}                  &      -\frac{m_{(2)}m_{(3)}}{\sqrt{m_{(1)}^2+m_{(2)}^2}| m_i|}                 &       \frac{\sqrt{m_{(1)}^2+m_{(2)}^2}}{| m_i|}
\end{pmatrix},
\end{equation*}
where $|m_i|=\sqrt{\sum_{i=1}^3 m_{(i)}^2}$, $\|m _ \mu\|=\sqrt{ m_{(0)}^2-|m_i |^2}$. And $\B _1 = \C_1 \C_2 \C_3$,
\begin{equation*}
\begin{aligned}
\C_1=&
\begin{pmatrix}
\frac{m_{(0)}}{\| m_\mu\|}       &      -\frac{| m_i|}{\| m_\mu\|}        &      0        &       0  \\
-\frac{| m_i|}{\| m_\mu\|}       &     \frac{m_{(0)}}{\| m_\mu\|}         &      0        &       0   \\
0               &       0                 &      1         &       0   \\
0               &      0                &      0                &      1
\end{pmatrix}, \\
\C_2=&
\begin{pmatrix}
1         &      0       &      0       &       0  \\
0         &      \frac{\sqrt{m_{(1)}^2+m_{(2)}^2}}{| m_i| }         &     0     &     \frac{m_{(3)}}{| m_i| }   \\
0         &       0           &      1        &       0   \\
0         &      -\frac{m_{(3)}}{| m_i| }                           &     0     &      \frac{\sqrt{m_{(1)}^2+m_{(2)}^2}}{| m_i| }
\end{pmatrix},\\
\C_3=&
\begin{pmatrix}
1         &      0       &      0       &       0  \\
0         &      \frac{m_{(1)}}{\sqrt{m_{(1)}^2+m_{(2)}^2 } }        &     \frac{m_{(2)}}{\sqrt{m_{(1)}^2+m_{(2)}^2 }}    &    0    \\
0         &      -\frac{m_{(2)}}{\sqrt{m_{(1)}^2+m_{(2)}^2 } }       &     \frac{m_{(1)}}{\sqrt{m_{(1)}^2+m_{(2)}^2 }}    &    0   \\
0         &      0       &     0     &      1
\end{pmatrix}.
\end{aligned}
\end{equation*}
Under this transformation, $c_{(i)}$ and $J_{(i)(j)}$ will also be changed under $\B_1$. Denote by $c_{(i)}^{(1)}$ and
$J_{(i)(j)}^{(1)}$ the respective new quantities. The following $SO(3,1)$ matrix $\B_2$ changes both $J_{(1)(3)}^{(1)}$ and
$J_{(2)(3)}^{(1)}$ to zero, and $J_{(1)(2)}^{(1)}$ to $|J^{(1)}|=\sqrt{(J ^{(1)} _{(1)(2)})^2+(J ^{(1)} _{(1)(3)})^2+(J ^{(1)} _{(2)(3)})^2}$
which is denoted by $J_{(1)(2)}^{(2)}$,
\begin{equation*}\label{B2}
\B_2=
\begin{pmatrix}
1          &             0                 &          0            &      0       \\
0          &           \frac{J ^{(1)} _{(1)(2)}}{|J^{(1)} _{13}|}       &   0      &    -\frac{J ^{(1)} _{(2)(3)}}{|J^{(1)} _{13}|}    \\
0          &           \frac{J ^{(1)} _{(1)(3)}J ^{(1)} _{(2)(3)}}{|J^{(1)} _{13}||J^{(1)}|}    &          \frac{|J^{(1)} _{13}|}{|J^{(1)}|}   &           \frac{J^{(1)} _{(1)(2)}J^{(1)} _{(1)(3)}}{|J^{(1)} _{13}||J^{(1)}|}  \\
0          &           \frac{J^{(1)} _{(2)(3)}}{|J^{(1)}|}    &           -\frac{J^{(1)} _{(1)(3)}}{|J^{(1)}|}   &           \frac{J^{(1)} _{(1)(2)}}{|J^{(1)}|}
\end{pmatrix},
\end{equation*}
where $|J^{(1)} _{13}|=\sqrt{|J^{(1)}|^2 - (J ^{(1)} _{(1)(3)})^2}$.
Also, $c_{(i)}^{(1)}$ will be changed, and we denote the corresponding new quantities by $c_{(i)}^{(2)}$. Now the following
$SO(3,1)$ matrix $\B_3$ changes $c_{(2)}^{(2)}$ to zero and preserves $J_{(1)(2)}^{(2)}$,
\begin{equation*} \label{B3}
\B_3=
\begin{pmatrix}
1          &             0                 &          0            &        0       \\
0          &           \frac{c_{(1)}^{(2)}}{\sqrt{{\big(c_{(1)}^{(2)}\big)}^2+{\big(c_{(2)}^{(2)}\big)}^2}}       &   \frac{c_{(2)}^{(2)}}{\sqrt{{\big(c_{(1)}^{(2)}\big)}^2+{\big(c_{(2)}^{(2)}\big)}^2}}       &          0    \\
0          &          - \frac{c_{(2)}^{(2)}}{\sqrt{{\big(c_{(1)}^{(2)}\big)}^2+{\big(c_{(2)}^{(2)}\big)}^2}}      &    \frac{c_{(1)}^{(2)}}{\sqrt{{\big(c_{(1)}^{(2)}\big)}^2+{\big(c_{(2)}^{(2)}\big)}^2}}
&          0  \\
0          &            0                 &          0            &          1
\end{pmatrix}.
\end{equation*}
Again $c_{(i)}^{(2)}$ and $J_{(i)(j)}^{(2)}$ will change to the new quantities which are denoted by $c_{(i)}^{(3)}$ and $J_{(i)(j)}^{(3)}$.

Thus the transformation $\B=\B_3 \B_2 \B_1$ changes the coordinates $\{y^\alpha \}$ to the center of AdS mass coordinates such that
(\ref{cond-CMT}) holds.

\mysection{Appendix C: Roots of $\det {\bf Q}$}

We compute explicitly four formal roots of the determinant $\det {\bf Q}$ of the energy-momentum endomorphism ${\bf Q}$ given by (\ref{Q}).
The equation $\det {\bf Q}=0$ is a quartic equation with the variable $E_0$.
Denote
\beQ
\begin{aligned}
\xi_1=&\sqrt{27 V^{12}+4 L^6 V^6-18{A^4} L^2 V^6-A^8 L^4+4 {A^{12}} },\\
\xi_2=&\sqrt[3]{2 L^6-9 {A^4}L^2+27 V^6+3 \sqrt{3} \xi_1},\\
\eta_1=&\sqrt{\frac{4 L^2}{3}+\frac{2}{3} 2^{2/3}{\xi_2}+\frac{4 \sqrt[3]{2}\left(L^4-3 {A^4}\right)}{3 {\xi_2}}},\\
\eta_2=&\sqrt{-\frac{16V^3}{{\eta_1}}+\frac{8 L^2}{3}-\frac{2}{3} 2^{2/3}{\xi_2}-\frac{4 \sqrt[3]{2} \left(L^4-3 {A^4}\right)}{3{\xi_2}}}.
\end{aligned}
\eeQ
If all of these are well-defined, then $\det {\bf Q}$ has four roots
\beQ
\frac{1}{2}\big(\pm \eta_1 \pm \eta_2\big).
\eeQ
In this case that $\det {\bf Q}\geq 0$ gives
\beQ
E_0 ^2 \geq \frac{1}{4}\big(\eta _1 +\eta _2\big)^2, \quad \mbox{or} \quad E_0 ^2 \leq \frac{1}{4}\big(\eta _1 -\eta _2\big)^2.
\eeQ

\end{document}